\documentclass[
 reprint,
 amsmath,amssymb,
 aps,
]{revtex4-1}

\usepackage{graphicx}
\usepackage{dcolumn}
\usepackage{bm}

\usepackage{soul}

\def\Vec#1{\boldsymbol #1}
\def\dps{\displaystyle}
\def\arrayret{\vspace{5pt} \\}

\begin{document}


\title{Torus knots as Hopfions}

\author{Michikazu Kobayashi$^{1}$, Muneto Nitta$^2$}
\affiliation{%
$^1$Department of Physics, Kyoto University, Oiwake-cho, Kitashirakawa, Sakyo-ku, Kyoto 606-8502, Japan, \\
$^2$Department of Physics, and Research and Education Center for Natural Sciences, Keio University, Hiyoshi 4-1-1, Yokohama, Kanagawa 223-8521, Japan
}%

\date{\today}

\begin{abstract}
We present a direct connection between torus knots and Hopfions by finding stable and static solutions of the extended Faddeev-Skyrme model with a 
ferromagnetic potential term.
$(P,Q)$--torus knots consisting of $|Q|$ sine-Gordon kink strings twisted $P/Q$ times into the poloidal cycle along the toroidal cycle on a toroidal domain wall 
carry the Hopf charge $PQ$, 
which demonstrates that Hopfions can be further classified according to torus knot type.
\end{abstract}


\maketitle


Around one and half centuries ago, Lord Kelvin proposed that atoms are made of vortex knots \cite{Kelvin}. Although this theory was rejected, his research lead Tait to devise the celebrated knot theory in mathematics. Since then, knot theory has become an important subject in topology.  
One useful construction of knots is the torus knot, in which 
a braid group on strings on a torus is manipulated.
Torus knots are characterized by two integers $P$ and $Q$ 
representing the number of string twists along the torus and the number of strings, respectively.   
Torus knots of vortices or line defects have been investigated broadly in physics \cite{various}:
fluid mechanics and plasma \cite{Moffatt:1969}, 
helium superfluid \cite{Ricca:1999}, 
Bose-Einstein condensates of ultracold atoms \cite{Proment:2012}, 
nonequilibrium systems \cite{Aranson:2002},
colloids \cite{Tkalec:2011}, optics \cite{Dennis:2012},
excited media \cite{Sutcliffe:2003gm}, 
quantum chromodynamics \cite{Buniy:2002yx}, and classical field theory \cite{Eto:2012vx}.

On the other hand, in high-energy physics, a field theoretical model admitting knot-like 
solitons, namely, the Faddeev-Skyrme (FS) model 
\cite{Faddeev:1975}, which 
is an $O(3)$ sigma model with four derivative (Skyrme) terms, was proposed.
Stable (un)knots were first constructed in 
Refs. \cite{Gladikowski:1996mb,Faddeev:1996zj}.
More generally, this model admits solitons having 
a topological charge, {\emph i.e.},
a Hopf charge classified by the homotopy group 
$\pi_3(S^2)\simeq {\bf Z}$, 
which are referred to as Hopfions \cite{Battye:1998pe,Radu:2008pp}. 
(Un)stable Hopfions have also been investigated in various physical systems such as exotic superconductors \cite{Babaev:2001zy}, 
ferromagnets \cite{Sutcliffe:2007vm}, 
and Bose-Einstein condensates \cite{Kawaguchi:2008xi}.
Stable Hopfions with higher Hopf charges 
in the FS model were numerically 
constructed \cite{Battye:1998pe},
and in particular the first non-trivial  knot structure 
appears in Hopfions with the Hopf charge 7  
\cite{Hietarinta:2000ci,Sutcliffe:2007ui}. 
Knot structures were also found in 
Hopfions with potential terms  \cite{Foster:2010zb,Harland:2013uk,Battye:2013xf}.
While it was conjectured \cite{Battye:1998pe} 
that all torus knots can be constructed as Hopfions, 
a precise connection between Hopfions and knots remains unclear. 

In our previous paper \cite{Kobayashi:2013bqa}, 
we considered the Faddeev-Skyrme model with  
a ferromagnetic potential term, that is, 
a potential term quadratic in the field \cite{Weidig:1998ii,Nitta:2012kk} 
admitting two discrete vacua and a domain wall 
with a $U(1)$ modulus 
interpolating between them \cite{Abraham:1992vb,Kudryavtsev:1997nw,Nitta:2012kk}. 
Hopfions can be constructed as 
twisted closed lump (baby Skyrmion) strings \cite{deVega:1977rk,Gladikowski:1996mb}, which are characterized by the number of twists $P$ and 
the lump (baby Skyrmion) charge $Q$, 
which takes a value in
$\pi_2(S^2) \simeq {\bf Z}$ 
of the constituent lumps. 
We found that the Hopfions are all in a toroidal shape, 
that is, toroidal domain walls, 
and that the $U(1)$ modulus of the domain wall 
is twisted $P$ and $Q$ times along the toroidal 
and poloidal cycles of the torus, respectively. 
The Hopf charge $C$ was found to be 
the product $C = P Q$, and consequently 
the Hopfions with the Hopf charge $C$ can be further classified according to two topological charges 
$P$ and $Q$. 
We explicitly constructed stable $(P,Q)$ Hopfion solutions 
with the Hopf charge $1 \leq C \leq 6$.
An immediate question arises. 
Aren't there any knot structures for Hopfions 
with higher Hopf charges in this model, 
unlike the conventional Hopfions?

In this Letter, we find a direct connection 
between Hopfions and torus knots  
in a different manner from \cite{Battye:1998pe}.
We show this by deforming 
the ferromagnetic Faddeev-Skyrme model  \cite{Nitta:2012kk,Kobayashi:2013bqa},
with adding a further potential term linear in the field,
considered in the original baby Skyrme model \cite{Piette:1994ug}. 
The total potential is well known in condensed matter physics as ferromagnets with two easy axes.
With this potential, we previously found 
in the case of $d=2+1$ dimensions that 
$|Q|$ sine-Gordon kinks appear on 
a domain wall \cite{Nitta:2012xq}
or a domain wall ring \cite{Kobayashi:2013ju}
with the lump charge $Q$. 
For the $(P,Q)$ Hopfions, 
$|Q|$ sine-Gordon kink strings appear 
on the toroidal domain wall, 
which are twisted $P/Q$ times into the poloidal cycle 
along the toroidal cycle 
with forming $(P,Q)$--torus knots. 
We numerically construct stable Hopfions in this model 
and find that the Hopf charge $C$ is 
the product $P Q$ of $(P,Q)$--torus knots. 
In other words, Hopfions with the Hopf charge $C$ can be further classified according to 
the torus knot type.  

We start with the Lagrangian density of the FS model:
\begin{align}
\begin{array}{c}
\dps \mathcal{L} = \frac{1}{2} \partial_\mu \Vec{n} \cdot \partial^\mu \Vec{n} - \kappa F^{\mu \nu} F_{\mu \nu} - V(\Vec{n}), \arrayret
\dps \Vec{n} \cdot \Vec{n} = 1, \quad F_{\mu \nu} = \Vec{n} \cdot \left(\partial_\mu \Vec{n} \times \partial_\nu \Vec{n} \right),
\end{array} \label{eq-FS-model}
\end{align}
where a unit three vector $\Vec{n} = (n_1(x), n_2(x), n_3(x))$ 
of scalar fields is characterized as a point in the $S^2$ target space.
In the original FS model with no potential, $V(\Vec{n}) = 0$, 
three-dimensional Hopfion structures are stabilized 
in addition to the vacuum state.
Next, we introduce the potential:
\begin{align}
\begin{array}{c}
\dps V(\Vec{n}) = V_1(\Vec{n}) + V_2(\Vec{n}), \arrayret
\dps V_1(\Vec{n}) = m^2 (1 - n_3) (1 + n_3), \quad V_2(\Vec{n}) = - \beta^2 n_1.
\end{array} \label{eq-FS-potential}
\end{align}
In the context of the baby Skyrme model in $d=2+1$, 
only $V_2$ \cite{Piette:1994ug} or $V_1$ 
\cite{Kudryavtsev:1997nw,Weidig:1998ii} 
was considered. 
Here, we consider both terms 
in the regime $m \gg \beta > 0$ \cite{Nitta:2012xq,Kobayashi:2013ju}. 
This potential is known in ferromagnets with two easy axes.
The potential $V_1$ allows two discrete vacua $n_3 = \pm 1$ 
and a domain wall interpolating between the vacua.
$V_2(\Vec{n})$ slightly shifts  these vacua to
${\Vec n}=(-\beta^2/2m,0,\pm \sqrt{1-\beta^4/4m^2})$,
but this shift is negligible in the regime of $\beta \ll m$.
It is, however,  
important for the domain wall solution.
Inside the domain wall where $n_3 = 0$, 
$\Vec{n}$ takes a value in $S^1$ within the $n_1$--$n_2$ plane, 
and $V_2(\Vec{n})$ chooses $n_1 = 1$ as the most stable state inside the domain wall.
In addition to the uniform $n_1 = 1$ solution on the domain wall, there also exists a sine-Gordon kink soliton solution constrained on the domain wall, where $\Vec{n}$ in $S^1$ is wound from one $n_1 = 1$ point to the other $n_1 = 1$ point through the $n_1 = -1$ point in the $n_1$--$n_2$ plane. In the $(3+1)$--dimensional space, the kink soliton is a $(1+1)$--dimensional string on the $(2+1)$--dimensional domain wall world-volume \cite{Nitta:2012xq}.

\begin{figure}[tbh]
\centering
\vspace{-6cm}
\includegraphics[width=1.0\linewidth]{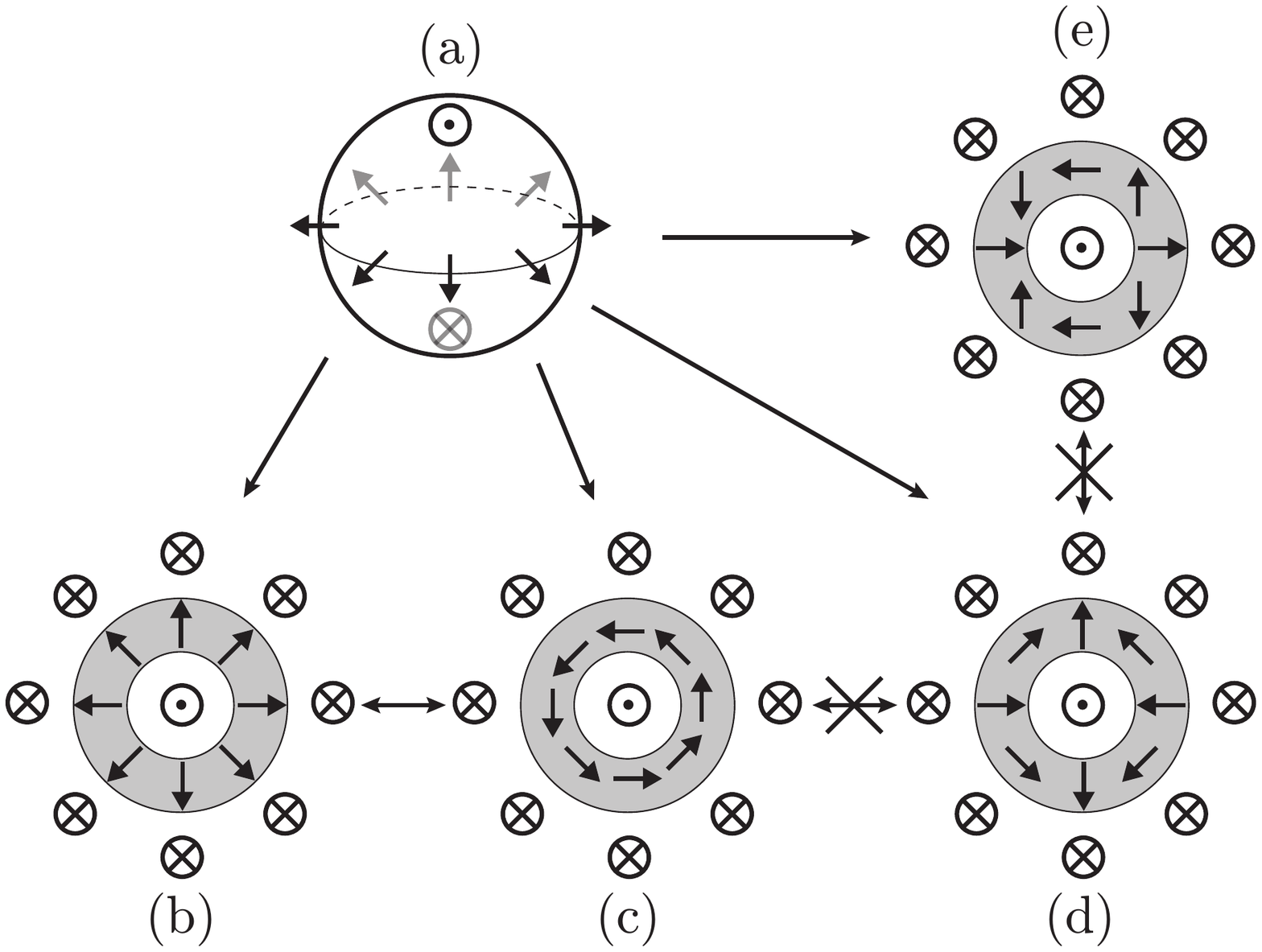}
\caption{\label{fig-baby-skyrmion} Examples of baby Skyrmion structure.
(a) $S^2$ target space of $\Vec{n}$. $n_3 = 1$ (north pole) and $n_3 = -1$ (south pole) are denoted by $\odot$ and $\otimes$, respectively. Directions on the equator ($n_3 = 0$) are denoted by arrows: $(n_1,n_2) = (1,0), (- 1,0), (0,1),$ and $(0,-1)$ for right, left, up, and down arrows, respectively. 
(b)--(e) Spatial configurations of $\Vec{n}$ for baby Skyrmions.
The gray shaded annuli denote regions in which $n_3 \sim 0$.
The topological charges defined in Eq.~\eqref{eq-baby-skyrmion-charge} of baby Skyrmions in (b)[(c)], (d), and (e) are $+1$, $-1$, and $+2$, respectively.}
\end{figure}

Here, we briefly recall the Hopfion structures in the original FS model without the potential term, starting from the baby Skyrmion (lump) string in the FS model. Baby Skyrmions can be constructed by mapping the two-dimensional space at the fixed boundary to the $S^2$ target space for $\Vec{n}$, as shown in Fig. \ref{fig-baby-skyrmion}(a). Fig. \ref{fig-baby-skyrmion}(b)--(e) shows typical spatial configurations of $\Vec{n}$ for baby Skyrmions. For all of the baby Skyrmions shown in these figures, the states at the center and the boundary are fixed to $n_3 = 1$ and $n_3 = -1$, respectively. Differences between these figures can be found in the configurations in the gray colored annuli areas, where $\Vec{n}$ is twisted once in the clockwise direction for Figs. \ref{fig-baby-skyrmion}(b) and \ref{fig-baby-skyrmion}(c), once in the anticlockwise direction for Fig.~\ref{fig-baby-skyrmion}(d), and twice in the clockwise direction for Fig.~\ref{fig-baby-skyrmion}(e) within the $n_1$--$n_2$ plane along the annulus. This number of twists counts the number of baby Skyrmions and is equivalent to the topological lump (baby Skyrmion) charge  
\begin{align}
Q = \frac{1}{8 \pi} \int d^2 x \: \epsilon^{\mu \nu} F_{\mu \nu}. \label{eq-baby-skyrmion-charge}
\end{align}
The charge $Q$ is $+1$ in Figs. \ref{fig-baby-skyrmion}(b) and \ref{fig-baby-skyrmion}(c), $-1$ in Fig. \ref{fig-baby-skyrmion}(d), and $+2$ in Fig. \ref{fig-baby-skyrmion}(e).
Baby Skyrmions, as shown in Figs. \ref{fig-baby-skyrmion}(b) and \ref{fig-baby-skyrmion}(c) can be continuously transformed by twisting $\Vec{n}$ around the $n_3$ direction, because these Skyrmions have the same topological charges $Q$.

\begin{figure}
\centering
\vspace{-2cm}
\includegraphics[width=1.0\linewidth]{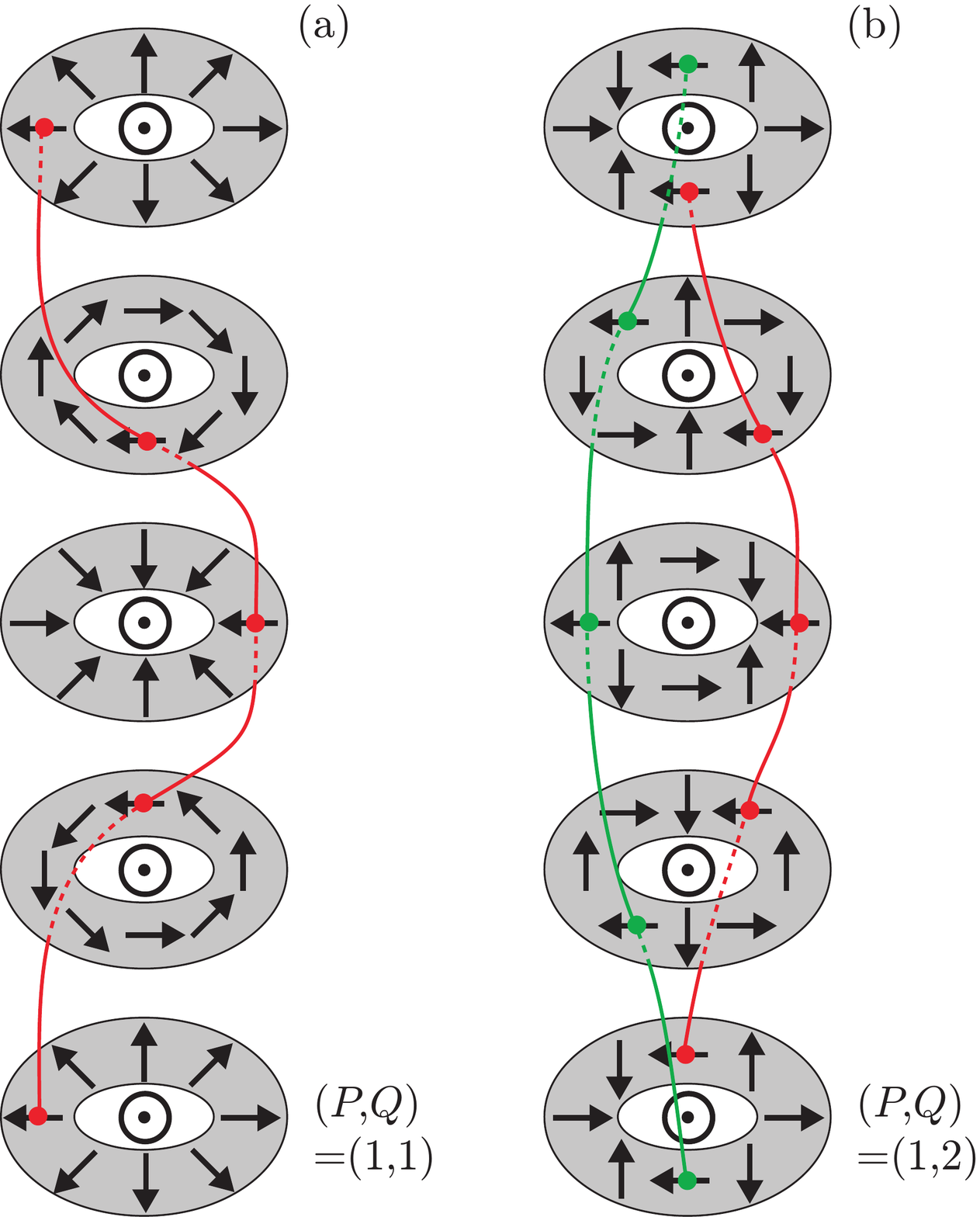}
\caption{\label{fig-cylinder} Twisted baby Skyrmion string.
The gray $n_3 = 0$ surfaces have cylindrical structures that separate the regions of $\odot$ and $\otimes$. 
Along the cylinders, there are sequences of baby Skyrmions with the charges of (a) $Q = 1$ and  (b) $Q = 2$. From the bottom baby Skyrmion to the top baby Skyrmion, $\Vec{n}$ rotates by $2 \pi P$ ($P = 1$) in the $n_1$--$n_2$ plane, defining the number of twists $P$.
The red and green lines indicate the locus of the $n_1 = -1$ state along the cylinder.}
\end{figure}

Next, we consider three-dimensional structures 
constructed from baby Skyrmions.
The simplest and non-trivial one is the twisted baby Skyrmion strings 
shown in Fig. \ref{fig-cylinder}.
There are sequences of baby Skyrmions along 
one direction (the $z$--axis) in real space, 
and each slice is made of baby Skyrmions with the lump charges 
$Q=1$ and $Q=2$  
in Fig.~\ref{fig-cylinder}(a) and \ref{fig-cylinder}(b), respectively. 
The number of twists $P$ in a segment of the twisted baby Skyrmion string 
is defined as the rotation angle $2 \pi P$ of $\Vec{n}$ 
in the $n_1$--$n_2$ plane in the target space 
from the bottom baby Skyrmion to the top baby Skyrmion
[$P = 1$ for both Figs. \ref{fig-cylinder}(a) and \ref{fig-cylinder}(b)].
In the figures, we further show the loci of $n_1 = -1$, 
which become kink soliton strings 
on a cylindrical domain wall 
when $V(\Vec{n}) \neq 0$.
The locus of $n_1 = -1$ is rotated clockwise along the axis by $2 \pi$ from the bottom to the top 
in Fig.~\ref{fig-cylinder}(a), and each of two loci of $n_1 = -1$ is rotated by $\pi$ in Fig.~\ref{fig-cylinder}(b). In general, when a twisted baby Skyrmion string of the lump charge $Q$
has $P$ twists, there exit $Q$ loci of $n_1 = -1$ rotated clockwise along the axis by $2 \pi P / Q$.

\begin{figure}[tbh]
\centering
\includegraphics[width=1.0\linewidth]{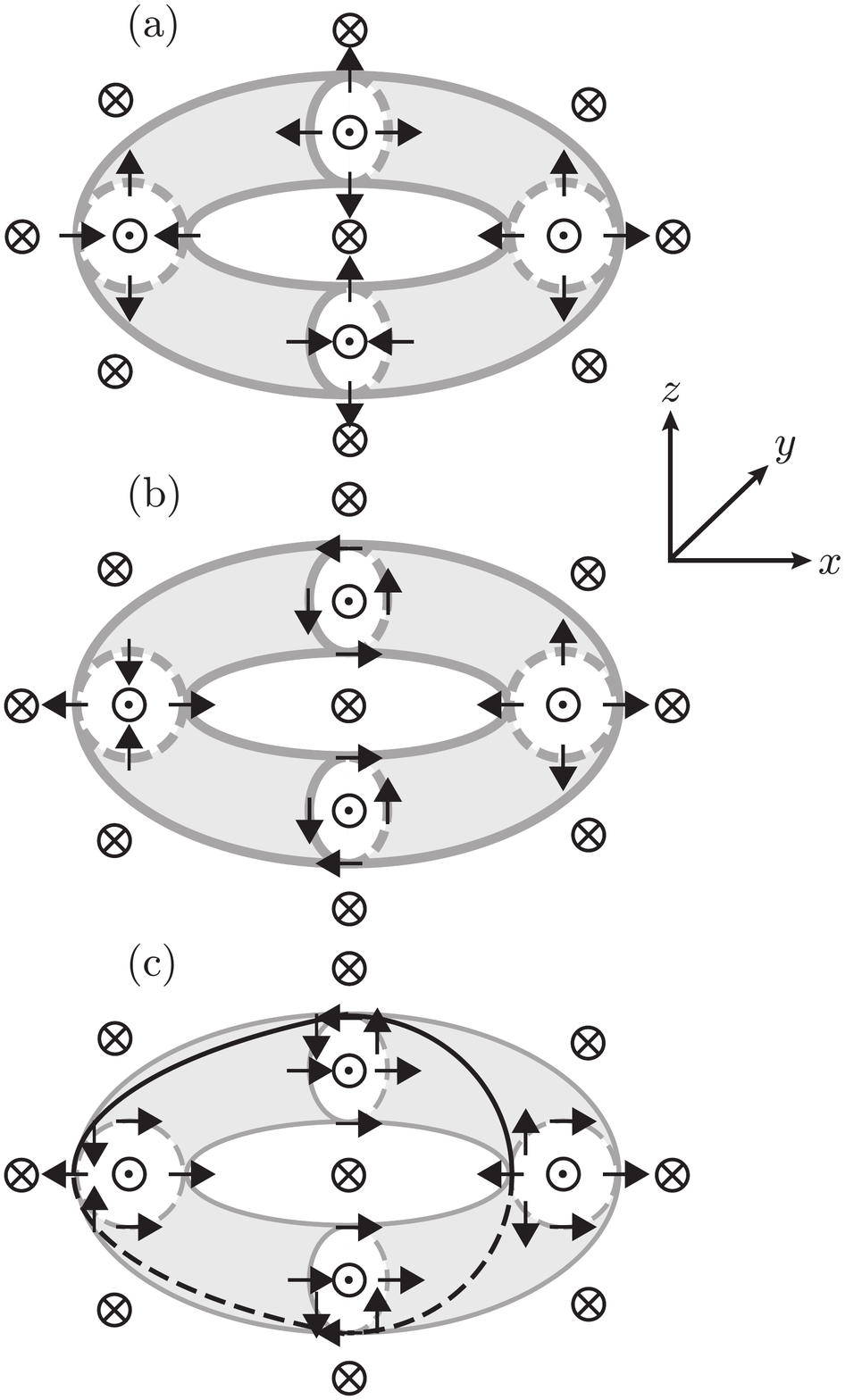}
\caption{\label{fig-hopfion} Untwisted (a) and twisted (b) baby Skyrmion rings.
The gray $n_3 = 0$ surfaces have torus structures that separate the regions of $\odot$ and $\otimes$. As the baby Skyrmion ring is rotated along the $z$--axis with from $0$ to $2 \pi$, $\Vec{n}$ on the torus does not twist in (a) and twists by $2 \pi$ in (b).
(c) Deformed twisted baby Skyrmion ring with $V(\Vec{n}) \neq 0$.
When $V(\Vec{n})$ is switched on, the $n_1 = 1$ ($\rightarrow$) state on the gray toroidal domain wall is stabilized, which makes the $n_1 = -1$ ($\leftarrow$) state the kink soliton string constrained on the toroidal domain wall. The thick line indicates the position of the kink soliton.}
\end{figure}

Next, we look at Hopfions.
Hopfions can be constructed as twisted closed baby Skyrmion strings 
by joining the tops and bottoms of twisted baby Skyrmion strings 
\cite{deVega:1977rk,Gladikowski:1996mb,Nitta:2012kk}, 
as shown in Fig.~\ref{fig-hopfion}. In baby Skyrmion rings, the surface defined by $n_3 = 0$ is in the form of a torus dividing the region of $n_3 = \pm 1$. On any section of the torus in a plane containing the $z$--axis, there is a pair of a baby Skyrmions with charge $Q(>0)$ and an anti-baby Skyrmion with charge $-Q$. In Fig.~\ref{fig-hopfion}(a), the number of twists is $P = 0$ for the constituent baby Skyrmion, and the configuration of $\Vec{n}$ does not change along the ring. In Fig.~\ref{fig-hopfion}(b), on the other hand, $\Vec{n}$ is rotated by $2 \pi$ in the $n_1$--$n_2$ plane in the target space along the ring with the number of twists  $P = 1$. In general, baby Skyrmion rings can be characterized by the topological lump charge $Q$ of the constituent baby Skyrmions and the number of twists  $P$ along the ring. The Hopf charge $C$ of $\pi_3(S^2)\simeq {\bf Z}$ defined by
\begin{align}
\begin{array}{c}
\dps C = \frac{1}{4 \pi^2} \int d^3x \: \epsilon^{\mu \nu \rho} F_{\mu \nu} A_\rho, \quad
\dps A_\mu = \frac{\epsilon_{ijk} n_i \partial_\mu n_j}{3 (1 + n_k)}
\end{array} \label{eq-hopfion-charge}
\end{align}
is equivalent to the product $PQ$ of the number of twists $P$ and the constituent lump charge $Q$. 
\begin{figure*}[htb]
\centering
\vspace{-0cm}
\includegraphics[width=0.75\linewidth]{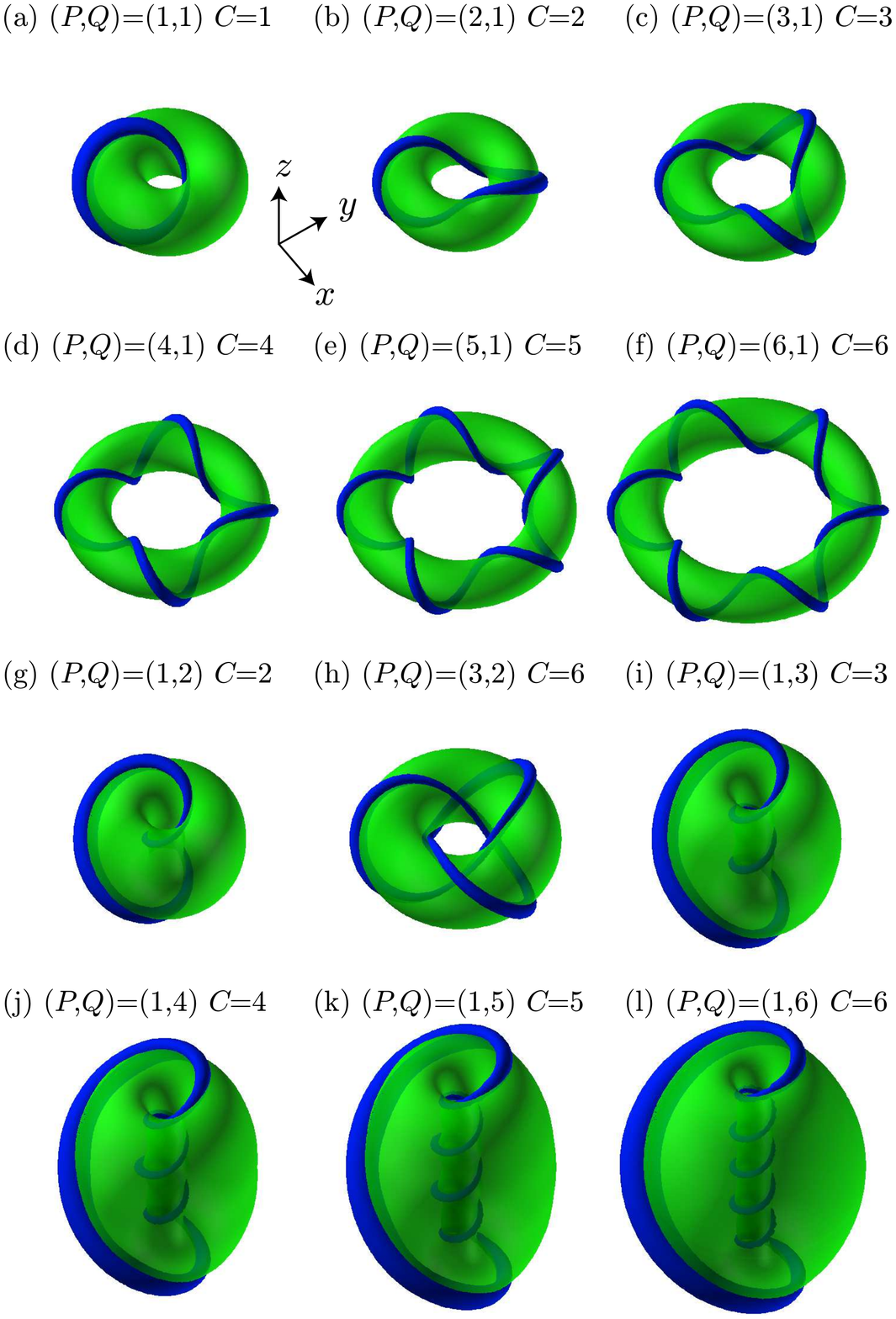}
\caption{\label{fig-torus-knot} Toroidal domain wall (transparent green surface) and kink soliton string (blue surface).
The Hopf charge $C$ is 1 in (a), 2 in (b) and (g), 3 in (c) and (i), 4 in (d) and (j), 5 in (e) and (k), and 6 in (f), (h), and (l).
The green surface is the isosurface of $n_3 = 0$.
The blue surface is isosurface of $n_1 = - 0.97$.
We fix $\beta^2 / m^2 = 0.01$, and $\kappa / m^2 = 1 \times 10^{-6}$.}
\end{figure*}
To show this, we promote configurations with 
the target space $S^2$ to 
those with $S^3$ by the Hopf map.
We introduce two complex scalar fields 
$\phi^T=(\phi_1,\phi_2)$ with the constraint of
$|\phi_1|^2+|\phi_2|^2=1$ which parametrizes $\phi$ by $S^3 \simeq SU(2)$.
The three-vector scalar fields $n_i$ 
can be written by the Hopf map 
\begin{align}
  n_i = \phi^\dagger \sigma_i \phi , \quad (i=1,2,3)
\label{eq:Hopf-map}
\end{align}
with the Pauli matrices $\sigma_i$.
We next consider an ansatz
\begin{align}
\begin{array}{c}
\displaystyle \phi = \begin{pmatrix} \cos R e^{- i P \phi} & \sin R e^{i Q \Theta} \end{pmatrix}, \vspace{5pt} \\
\displaystyle R = \cos^{-1}\{\sin f(r) \sin\theta\}, \vspace{5pt} \\
\displaystyle \Theta = - \tan^{-1} \bigg\{\frac{\sin f(r) \cos\theta}{- \cos f(r)}\bigg\}
\end{array}
\label{eq:ansatz}
\end{align}
with the polar coordinate $(r, \theta, \phi)$.
Here, a monotonically increasing function $f(r)$ satisfies
the boundary condition 
\begin{align}
f(r \to 0) \to 0, \quad
f(r \to \infty) \to \pi.
\end{align}
From the Hopf map in Eq.~(\ref{eq:Hopf-map}), 
we have
\begin{align}
\begin{array}{c}
\displaystyle \Vec{n} = \begin{pmatrix} \sin(2 R) \cos\Phi & \sin(2 R) \sin\Phi & \cos(2 R) \end{pmatrix}, \vspace{5pt} \\
\displaystyle \Phi = P \phi + Q \Theta.
\end{array}
\label{eq:Hopfion}
\end{align}
The configuration given in Eq.~\eqref{eq:Hopfion} 
is isomorphic to a torus knot with $(P, Q)$ and its Hopf charge $C$ can be obtained 
through the Hopf map in Eq.~(\ref{eq:Hopf-map})
from the Skyrme charge $\pi_3(S^3) \simeq {\bf Z}$ 
of the fields $\phi$ in Eq.~\eqref{eq:ansatz}:
\begin{align}
\begin{split}
C &:= \frac{1}{2 \pi^2} \int d^3x\: \epsilon^{abcd} 
s_a \partial_1 s_b \partial_2 s_c \partial_3 s_d \\
&= \frac{P Q}{\pi} \int_0^\infty dr\: \frac{d}{dr} \left\{ f(r) - \sin f(r) \cos f(r) \right\} \\
&= P Q,
\end{split}
\end{align}
with $\phi_1 = s_1 + i s_2$ and $\phi_2 = s_3 + i s_4$ 
($a,b,c,d=1,2,3,4$).

We now consider the potential term $V(\Vec{n})$, 
which allows the vacua $n_3 \sim \pm 1$. For baby Skyrmions in $d=2+1$, 
rings defined by $n_3 = 0$ become domain wall rings 
by the potential $V_1$. Furthermore, on the domain wall rings, the points of $n_1 = 1$ are stabilized and the points of $n_1 = - 1$  become sine-Gordon solitons 
by the potential $V_2$. As a result, $|Q|$ sine-Gordon solitons appear on domain wall rings as baby Skyrmions with the lump charge $Q$ in $d=2+1$ \cite{Kobayashi:2013ju}.  
For the Hopfions in $d=3+1$, the toroidal surface defined by $n_3=0$ 
becomes a domain wall separating the two vacua $n_3 = \pm 1$, as shown in Fig.~\ref{fig-hopfion}(c). Furthermore, $|Q|$ sine-Gordon soliton strings appear on the toroidal domain wall.
Along the longitude of the torus of the domain wall, the kink soliton strings are rotated by $2 \pi P / Q$, forming a $(P,Q)$--torus knot.
Therefore, the Hopfions with the charge $C(=PQ)$ in the original FS model correspond to $(P, Q)$--torus knots of the kink soliton strings on the toroidal domain wall in the FS model with $V(\Vec{n}) \neq 0$. In other words, Hopfions are further classified into topologically distinct $(P,Q)$ torus knots with fixed $C = PQ$.

We numerically construct solutions of $(P,Q)$--torus knots 
by solving the Euler-Lagrange equation
\begin{align}
\frac{\delta \mathcal L}{\delta \Vec{n}} = \frac{\partial \mathcal L}{\partial \Vec{n}} - \partial_\mu \frac{\partial \mathcal L}{\partial (\partial_\mu \Vec{n})} = 0 \label{eq-stationary-FS}
\end{align}
in the relaxation method. 
We introduce a relaxation time $\tau$ and the dependence of 
$\Vec{n}$ on $\tau$, and calculate the large-$\tau$ behavior for the equation:
\begin{align}
\frac{\partial \Vec{n}}{\partial \tau} = - \frac{\delta \mathcal L}{\delta \Vec{n}}.
\end{align}
The initial configurations of $\Vec{n}$ at $\tau = 0$ are given by Eq. \eqref{eq:Hopfion}, 
the configuration of which is isomorphic to a Hopfion with the charge $C = P Q$. Furthermore, in this configuration, the isosurface of $n_3 =0$ takes a torus configuration, and loci of $n_1 = -1$ on the torus take the form of 
a $(P,Q)$--torus knot.
As a result, we can obtain the solution 
for the $(P, Q)$--torus knot of kink solitons on the toroidal domain wall.

We numerically search the stable solution for the Hopf charge $1 \leq C \leq 6$.
Fig. \ref{fig-torus-knot}(a)--(l) shows different stable solutions of kink soliton string with $(P, Q) = (1,1)$, $(2,1)$, $(3,1)$, $(4,1)$, $(5,1)$, $(6,1)$, $(1,2)$, $(3,2)$, $(1,3)$, $(1,4)$, $(1,5)$, and $(1,6)$, respectively, and we find no stable solution for $(P, Q) = (2,2)$ or $(2,3)$.
While all torus knots except for Fig. \ref{fig-torus-knot}(h) are topologically equivalent to the trivial knot, the torus knot for Fig. \ref{fig-torus-knot}(h) is topologically equivalent to the trefoil knot.


All configurations shown in Fig.~\ref{fig-torus-knot} are located at stationary and minimal points of the Lagrangian $L = \int d^3x\: \mathcal L$. In the original FS model without the potential term, different states with the same Hopf charge $C$ are topologically equivalent and can be continuously deformed to each other under continuous changes of $\Vec{n}$.
In the case with $V(\Vec{n}) \neq 0$, configurations with different sets of $(P, Q)$ for torus knots of soliton strings are topologically distinct due to energy barriers between the knots, 
even when they have the same Hopf charge $C$.
Therefore, the configurations can be classified by the two integers $(P, Q)$ for torus knots rather than the Hopf charge $C$. 
Some of independent $(P,Q)$ Hopfions are stable, 
topologically distinct and energetically separated 
by the potential barrier even for the same Hopf number $C=PQ$.

In conclusion, we have investigated the extended FS model with the potential term given in Eq.~\eqref{eq-FS-potential}. In this model, the Hopfions in the original FS model can be expressed as the $(P, Q)$--torus knots of kink soliton strings of $n_1 = -1$ on the toroidal domain walls defined by $n_3=0$ interpolating between the two vacua $n_3 = \pm 1$. The toroidal domain walls are constructed as twisted rings of constituent lumps, and the two integers $P$ and $Q$ for the torus knots of kink soliton strings are the number of twists of $\Vec{n}$ along the ring and the topological lump charge of the constituent lumps, respectively. The product $PQ$ of the number of twists $P$ and the constituent lump charge $Q$ is equivalent to the Hopf charge $C$ defined in Eq.~\eqref{eq-hopfion-charge}. The Hopfions with the Hopf charge $C$ are further classified according to the different types of $(P, Q)$ torus knots with fixed $C = PQ$.
Some of them are stable, topological distinct, and energetically separated by the potential barrier.


\begin{thebibliography}{100}
\bibitem{Kelvin}
Sir William Thomson (Lord Kelvin), ``On Vortex Atoms," 
Proc. Roy. Soc. Edinburgh {\bf 6}, 94 E05 (1867).

\bibitem{various}
H.~K.~Moffat, 
``The energy spectrum of knots and links,"
Nature {\bf 347}, 367 E69 (1990); 
V.~Katritch, J.~Bednar, D.~Michoud, R.~G.~Scharein, 
J.~Dubochet and A.~Stasiak, 
``Geometry and physics of knots,"
Nature {\bf 384}, 142 - 145 (1996)

\bibitem{Moffatt:1969}
L.~Woltjer, Proc.\ Nat.\ Acad.\ Sci., {\bf 44}, 489 (1958);
H.~K.~Moffat, 
``The degree of knottedness of tangled vortex lines,"
J.\ Fluid\ Mech.\ {\bf 35}, 117-129  (1969).

\bibitem{Ricca:1999}
R.~L.~Ricca, D.~C.~Samuels,
and C.~F.~Barenghi,
``Evolution of vortex knots,"
J.\ Fluid\ Mech.\ {\bf 391}, 29-44 (1999).

\bibitem{Proment:2012}
D.~Proment, M.~Onorato, and C.~F.~Barenghi,
``Vortex knots in a Bose-Einstein condensate,"  
Phys.\ Rev.\ E {\bf 85}, 036306 (2012).

\bibitem{Aranson:2002}
I.~S.~Aranson and L.~Kramer, 
``The world of the complex Ginzburg-Landau equation,"
Rev.\ Mod.\ Phys.\ {\bf 74}, 99 E43 (2002) 

\bibitem{Tkalec:2011}
U.~Tkalec, M.~Ravnik, S.~\v{C}opar, S.~\v{Z}umer, and I.~Mu\v{s}evi\v{c},
``Reconfigurable Knots and Links in Chiral Nematic Colloids,"
Science {\bf 333}, 62-65 (2011).

\bibitem{Dennis:2012}
M.~R.~Dennis, R.~P.~King, B.~Jack, K.~O'Holleran, and M.~J.~Padgett, 
``Isolated optical vortex knots,"
Nature Physics {\bf 6}, 118 - 121 (2010); 
K.~O'Holleran, M.~R.~Dennis, and M.~J.~Padgett,
``Topology of Light's Darkness,"
Phys.\ Rev.\ Lett.\ {\bf 102}, 143902 (2009);
K.~O'Holleran, M.~R.~Dennis, F.~Flossmann, and M.~J.~Padgett, 
``Fractality of Light's Darkness,"
Phys.\ Rev.\ Lett. {\bf 100}, 053902 (2008); 
J.~Leach, M.~R.~Dennis, J.~Courtial and M.~J.~Padgett, 
``Laser beams:  Knotted threads of darkness,"
Nature {\bf 432}, 165 (2004);
W.~T.~M.~Irvine and D.~Bouwmeester
``Linked and knotted beams of light,"
 Nature Physics {\bf 4}, 817 (2008).

\bibitem{Sutcliffe:2003gm} 
A. T. Winfree,
``Persistent tangled vortex rings in generic excitable media,"
Nature {\bf 371}, 233 - 236 (1994); 
  P.~Sutcliffe and A.~Winfree,
  ``On the stability of knots in excitable media,'' 
Phys.\ Rev.\ E {\bf 68}, 016218 (2003) [nlin/0305025 [nlin-ps]].  



\bibitem{Buniy:2002yx} 
  R.~V.~Buniy and T.~W.~Kephart,
  ``A Model of glueballs,''  Phys.\ Lett.\ B {\bf 576}, 127 (2003)  [hep-ph/0209339].  

\bibitem{Eto:2012vx} 
  M.~Eto and S.~B.~Gudnason,
  ``Knotted domain strings,''  arXiv:1212.0702 [hep-th].  

\bibitem{Faddeev:1975}
L.~D.~Faddeev, Princeton preprint IAS-75-QS70.

\bibitem{Gladikowski:1996mb} 
  J.~Gladikowski and M.~Hellmund,
  ``Static solitons with nonzero Hopf number,''  Phys.\ Rev.\ D {\bf 56}, 5194 (1997)  [hep-th/9609035].  

\bibitem{Faddeev:1996zj} 
  L.~D.~Faddeev and A.~J.~Niemi,
  ``Knots and particles,''  Nature {\bf 387}, 58 (1997)  [hep-th/9610193].  

\bibitem{Battye:1998pe} 
  R.~A.~Battye and P.~M.~Sutcliffe,
  ``Knots as stable soliton solutions in a three-dimensional classical field theory.,''  Phys.\ Rev.\ Lett.\  {\bf 81}, 4798 (1998)  [hep-th/9808129];  
  R.~A.~Battye and P.~Sutcliffe,
  ``Solitons, links and knots,''  Proc.\ Roy.\ Soc.\ Lond.\ A {\bf 455}, 4305 (1999)  [hep-th/9811077].  

\bibitem{Radu:2008pp} 
  E.~Radu and M.~S.~Volkov,
  ``Existence of stationary, non-radiating ring solitons in field theory: knots and vortons,''  Phys.\ Rept.\  {\bf 468}, 101 (2008)  [arXiv:0804.1357 [hep-th]].  

\bibitem{Babaev:2001zy} 
  E.~Babaev, L.~D.~Faddeev and A.~J.~Niemi,
  ``Hidden symmetry and knot solitons in a charged two-condensate Bose system,''  Phys.\ Rev.\ B {\bf 65}, 100512 (2002)  [cond-mat/0106152 [cond-mat.supr-con]]; 
  E.~Babaev,
  ``Knotted solitons in triplet superconductors,''  Phys.\ Rev.\ Lett.\  {\bf 88}, 177002 (2002)  [cond-mat/0106360].  

\bibitem{Sutcliffe:2007vm} 
  P.~Sutcliffe,
  ``Vortex rings in ferromagnets,''  Phys.\ Rev.\ B {\bf 76}, 184439 (2007)  [arXiv:0707.1383 [cond-mat.mes-hall]].  

\bibitem{Kawaguchi:2008xi} 
  Y.~Kawaguchi, M.~Nitta and M.~Ueda,
  ``Knots in a Spinor Bose-Einstein Condensate,''  Phys.\ Rev.\ Lett.\  {\bf 100}, 180403 (2008)  [Erratum-ibid.\  {\bf 101}, 029902 (2008)]  [arXiv:0802.1968 [cond-mat.other]].  

\bibitem{Hietarinta:2000ci} 
  J.~Hietarinta and P.~Salo,
  ``Ground state in the Faddeev-Skyrme model,''  Phys.\ Rev.\ D {\bf 62}, 081701 (2000).  

\bibitem{Sutcliffe:2007ui} 
  P.~Sutcliffe,
  ``Knots in the Skyrme-Faddeev model,''  Proc.\ Roy.\ Soc.\ Lond.\ A {\bf 463}, 3001 (2007)  [arXiv:0705.1468 [hep-th]].  

\bibitem{Foster:2010zb} 
  D.~Foster,
  ``Massive Hopfions,''  Phys.\ Rev.\ D {\bf 83}, 085026 (2011)  [arXiv:1012.2595 [hep-th]].  

\bibitem{Harland:2013uk} 
  D.~Harland, J.~Jaykka, Y.~Shnir and M.~Speight,
  ``Isospinning hopfions,''  arXiv:1301.2923 [hep-th].  

\bibitem{Battye:2013xf} 
  R.~A.~Battye and M.~Haberichter,
  ``Classically Isospinning Hopf Solitons,''  arXiv:1301.6803 [hep-th].  

\bibitem{Kobayashi:2013bqa} 
  M.~Kobayashi and M.~Nitta,
  ``Toroidal domain walls as Hopfions,''  arXiv:1304.4737 [hep-th].  

\bibitem{Weidig:1998ii}
  T.~Weidig,
  ``The baby Skyrme models and their multi-skyrmions,''
Nonlinearity {\bf 12}, 1489-1503 (1999). 

\bibitem{Nitta:2012kk} 
  M.~Nitta,
  ``Knots from wall--anti-wall annihilations with stretched strings,''  Phys.\ Rev.\ D {\bf 85}, 121701 (2012)  [arXiv:1205.2443 [hep-th]].  

\bibitem{Abraham:1992vb} 
  E.~R.~C.~Abraham and P.~K.~Townsend,
  ``Q kinks,''  
Phys.\ Lett.\ B {\bf 291}, 85 (1992);  
  ``More on Q kinks: A (1+1)-dimensional analog of dyons,''  
Phys.\ Lett.\ B {\bf 295}, 225 (1992);  
  M.~Arai, M.~Naganuma, M.~Nitta and N.~Sakai,
  ``Manifest supersymmetry for BPS walls in N=2 nonlinear sigma models,''  Nucl.\ Phys.\ B {\bf 652}, 35 (2003)  [hep-th/0211103];  
  ``BPS wall in N=2 SUSY nonlinear sigma model with Eguchi-Hanson manifold,''  In *Arai, A. (ed.) et al.: A garden of quanta, 2003, 299-325  [hep-th/0302028].  

\bibitem{Kudryavtsev:1997nw}
  A.~E.~Kudryavtsev, B.~M.~A.~Piette and W.~J.~Zakrzewski,
  ``Skyrmions and domain walls in (2+1) dimensions,''
  Nonlinearity {\bf 11}, 783 (1998);
  D.~Harland and R.~S.~Ward,
  ``Walls and chains of planar skyrmions,''
  Phys.\ Rev.\  D {\bf 77}, 045009 (2008).

\bibitem{deVega:1977rk} 
  H.~J.~de Vega,
  ``Closed Vortices and the HOPF Index in Classical Field Theory,''  
Phys.\ Rev.\ D {\bf 18}, 2945 (1978);  
  A.~Kundu and Y.~P.~Rybakov,
  ``Closed Vortex Type Solitons With Hopf Index,''  
J.\ Phys.\ A A {\bf 15}, 269 (1982).  

\bibitem{Piette:1994ug}
  B.~M.~A.~Piette, B.~J.~Schroers and W.~J.~Zakrzewski,
  ``Multi - Solitons In A Two-Dimensional Skyrme Model,''
  Z.\ Phys.\  C {\bf 65}, 165 (1995);
  ``Dynamics of baby skyrmions,''
  Nucl.\ Phys.\  B {\bf 439}, 205 (1995).

\bibitem{Nitta:2012xq} 
  M.~Nitta,
  ``Josephson vortices and the Atiyah-Manton construction,''  Phys.\ Rev.\ D {\bf 86}, 125004 (2012)  [arXiv:1207.6958 [hep-th]];  
  M.~Nitta,
  ``Matryoshka Skyrmions,''  Nucl.\  Phys.\ B {\bf 872}, 62 (2013)  [arXiv:1211.4916 [hep-th]];  
  M.~Nitta,
  ``Instantons confined by monopole strings,''  Phys.\  Rev.\ D {\bf 87}, 066008 (2013)  [arXiv:1301.3268 [hep-th]].  

\bibitem{Kobayashi:2013ju} 
  M.~Kobayashi and M.~Nitta,
  ``Jewels on a wall ring,''  Phys.\  Rev.\ D {\bf 87}, 085003 (2013)  [arXiv:1302.0989 [hep-th]].  



\end{thebibliography}

%
We would like to thank the organizers of the ``Quantized Flux in Tightly Knotted and Linked Systems" conference held December 3--7, 2012 at Isaac Newton Institute for Mathematical Sciences, where the present study was initiated. The present study was supported in part by Grants-in-Aid for Scientific Research (Grants No. 22740219 (M.K.) and No. 23740198 and 25400268 (M.N.)) and the M.N. was also supported by a Grant-in-Aid for Scientific Research on Innovative Areas (``Topological Quantum Phenomena'') (No. 23103515 and 25103720) from the Ministry of Education, Culture, Sports, Science and Technology (MEXT) of Japan.
MK thanks the Supercomputer Center, the Institute for Solid State Physics, the University of Tokyo for the use of the facilities.

\end{document}